\documentclass[twocolumn,prl,nofootinbib,superscriptaddress,showpacs]{revtex4-1}
\usepackage{amsmath}
\usepackage{graphicx}

\newcommand{\be}{\begin{eqnarray}}
\newcommand{\ee}{\end{eqnarray}}
\newcommand{\forget}[1]{}

\begin{document}

\title{Nonlocal setting and outcome information for violation of Bell's inequality}

\author{Marcin Paw\l owski}
\affiliation{Institute for Quantum Optics and Quantum Information, Austrian Academy of
Sciences, Boltzmanngasse 3, A-1090 Vienna, Austria}
\affiliation{Institute of Theoretical Physics and Astrophysics, University of Gda\'nsk,
ul. Wita Stwosza 57, PL-80-952 Gda\'nsk, Poland}

\author{Johannes Kofler}
\affiliation{Institute for Quantum Optics and Quantum Information, Austrian Academy of
Sciences, Boltzmanngasse 3, A-1090 Vienna, Austria}
\affiliation{Faculty of Physics, University of Vienna, Boltzmanngasse 5, A-1090 Vienna,
Austria}

\author{Tomasz Paterek}
\affiliation{Institute for Quantum Optics and Quantum Information, Austrian Academy of
Sciences, Boltzmanngasse 3, A-1090 Vienna, Austria}
\affiliation{Centre for Quantum Technologies, National
University of Singapore, 3 Science Drive 2, 117542 Singapore, Singapore}

\author{Michael Seevinck}
\affiliation{Institute for Mathematics, Astrophysics and Particle Physics, Faculty of Science \& Centre for the History of Philosophy and Science, Faculty of Philosophy,  Radboud University Nijmegen, the Netherlands}
\affiliation{Institute for Theoretical Physics, Utrecht University,
P.O.Box 80.195, 3508 TD  Utrecht, the Netherlands.}

\author{{\v C}aslav Brukner}
\affiliation{Institute for Quantum Optics and Quantum Information, Austrian Academy of
Sciences, Boltzmanngasse 3, A-1090 Vienna, Austria}
\affiliation{Faculty of Physics, University of Vienna, Boltzmanngasse 5, A-1090 Vienna,
Austria}

\begin{abstract}
Bell's theorem is a no-go theorem stating that quantum mechanics cannot be reproduced by
a physical theory based on realism, freedom to choose experimental settings and two locality
conditions: setting (SI) and outcome (OI) independence. We provide a novel analysis of what it takes to violate Bell's inequality within the framework
in which both realism and freedom of choice are assumed, by showing  that it is impossible to model a
violation without having information in one laboratory about \textit{both}
the setting and the outcome at the distant one. While it is possible
that outcome information can be revealed from shared hidden variables, the assumed experimenter's freedom to choose the
settings forces that setting information \textit{must be} non-locally transferred, even when the SI condition is obeyed.
The sufficient amount of transmitted information about the setting to violate the
CHSH inequality up to its quantum mechanical maximum is 0.736~bits.
\end{abstract}

\pacs{03.65.Ta, 03.65.Ud}

\maketitle

Bell's inequalities are certain constraints on correlations between
space-like separated measurements which are satisfied in any local realistic
theory~\cite{BELL}. The inequalities are violated by quantum predictions for
some entangled states. The usual set of assumptions invoked in the
derivation of Bell's inequalities are \textit{realism}, the experimenter's
\textit{freedom} to choose the measurement settings (\textquotedblleft
freedom of choice\textquotedblright ), and two \textit{locality} conditions:
setting independence (SI) and outcome independence (OI)~\cite{Jarr,Shim,See}.
Maintaining realism and freedom of choice thus necessitates an exchange
of information between distant measurement stations that defies locality
so as to violate one (or both) of the independence conditions. What
kind of information and which amount of it has to be transferred between the
stations to model the violation of Bell's inequalities? Is it information
about the distant outcome, or about the distant setting, or about both?
These questions will be addressed in the present paper from an information theoretical perspective,
thereby providing a novel analysis of what it takes to violate Bell's inequality.

While we see other possibilities than maintaining realism and freedom of choice and introducing non-local actions to interpret the implications
of Bell's theorem, we acknowledge the importance of exploring alternative descriptions to deepen our understanding
of the foundations of quantum theory. In addition to fundamental reasons, answering the above questions is important
in quantum information, such as e.g. in quantum communication complexity problems~\cite{CCP}.
The question how much setting and/or outcome information needs to be exchanged in a Bell experiment for a
given degree of violation is relevant for quantifying the classical
resources required to simulate quantum efficiency in these problems.

In this work, we assume realism and freedom of choice
and study non-local hidden-variable models with one-way communication between two separated observers,
conventionally called Alice and Bob.
Clearly, communicating a distant setting allows simulation of violation of Bell's inequality.
For example, we could let Bob's outcome be determined by a shared hidden variable and we could have Alice's outcome depend
not only on her local setting and the shared hidden variable, but also on Bob's setting.
In this way, any set of correlations could be modeled as at Alice's location there is information about
all outcomes and settings that are involved.

Here we show that information about a distant setting
and outcome is not only sufficient to simulate violation
of Bell's inequality, but it is also necessary. This is to
be contrasted with the well-known condition \cite{Shim,See} that
either OI or SI for (hidden-variable) conditional probabilities
\textit{can still be obeyed} in models giving a violation of
Bell's inequality. Note that it is not a contradiction that
for a violation of Bell's inequality (I) both distant setting
and outcome information must be locally available
and (II) OI or SI can still be fulfilled. This is because
(I) and (II) refer to different notions. For instance, OI
is obeyed when the conditional (hidden variable) probability
for Alice's outcome does not \textit{explicitly} depend on
Bob's outcome. However, the information about Bob's
outcome can still be \textit{implicitly} contained in the shared
hidden variable. This allows for a novel analysis of what it
takes to violate Bell's inequality.

%

Furthermore we show, while it is possible that information about
the distant outcome can be read from the hidden-variables received from the
source, the information about the setting \textit{must} be non-locally
transmitted, implicitly or explicitly,
in any model where the experimenters are free to choose their settings.
We are able to trace
this asymmetry between setting and outcome information to the freedom of the experimenters to choose their settings.
We furthermore apply our analysis to the non-local hidden-variable models of Toner and Bacon \cite{TB}, Leggett \cite%
{LEGGETT} and Bohm \cite{BOHM}. Finally, we
show that the sufficient amount of transmitted information about the setting
to violate the CHSH inequality up to its quantum mechanical maximum  is 0.736~bits.

We begin with the usual formal definitions of the assumptions of Bell's theorem.
In our notation $a$ and $b$ stand for the measurement settings chosen by the
two distant experimenters Alice and Bob,
respectively; $A$ and $B$ denote their respective measurement outcomes, and $%
\lambda$ denotes a set of hidden variables.

(i)  For stochastic (probabilistic) hidden-variable theories the assumption of \textit{realism} dictates that the hidden variable $\lambda$ specifies joint
(non-negative, properly normalized) probabilities  $P(A_{1,1},A_{1,2},A_{2,1},..;B_{1,1},B_{1,2},B_{2,1},..|\lambda)$,
where e.g. the result $A_{1,2}$ of Alice for her setting choice 1 can depend on some
non-local parameter 2, typically Bob's setting choice. The conditional probabilities $P(A,B|a,b,\lambda)$ that will be used in this paper are then obtained as marginals of these joint ones.

(ii) \textit{Setting Independence} (SI), often also called Parameter
Independence \cite{Shim}, is the part of the locality condition which
prohibits the conditional dependence of the probability to obtain the outcome in
one laboratory on the
choice of the setting at the other one:
 $P(A|a,b,\lambda)=P(A|a,\lambda)$, and analogous
for $P(B|\cdot)$. Similarly, under \textit{Outcome Independence}
(OI), Alice's probability to obtain her outcome does not conditionally depend on
Bob's outcome and vice versa: $P(A|a,b,B,\lambda)=P(A|a,b,\lambda)$,
again analogous for $P(B|\cdot)$.
The conjunction of these two conditions is equivalent to Bell's condition of
Local Causality~\cite{bell,Jarr,Shim,See}:
$P(A,B|a,b,\lambda)=P(A|a,\lambda)P(B|b,\lambda)$.
The latter condition allows to define the joint probabilities from (i) as
$p(A_1,A_2|\lambda) p(B_1,B_2|\lambda)$, where e.g. $A_1$ is the
result of Alice for her setting choice $1$ \cite{Fine}. Note that fulfillment of SI
does \textit{not} imply that $P(A|a,b,B,\lambda)$ equals $P(A|a,B,\lambda)$. This shows that SI does
not entail \textit{complete} independence from the distant setting and that different criteria should be used.

(iii) The experimenter's \textit{freedom of choice} to choose the measurement setting
 imposes that the selected
measurement setting is statistically independent of the hidden variables
sent by the source (even in a deterministic model) \cite{BELL_FREE,PRA_FREE}.
In terms of the (Shannon) mutual information this
assumption is expressed as $I(\lambda :a)=I(\lambda :b)=0$.
As we will show, the assumption of freedom of choice is responsible for the
\emph{fundamental asymmetry} between settings and outcomes because it
guarantees that the settings, contrary to the outcomes, are to be considered
as independent variables.

Under these three assumptions the Clauser-Horne-Shimony-Holt (CHSH) inequality~\cite%
{CHSH} must be obeyed:
\begin{equation}
\frac{1}{4}\sum_{a,b=0}^{1}P(A\oplus B=ab|a,b)\leq \frac{3}{4},  \label{CHSH}
\end{equation}
with $\oplus$ denoting addition modulo 2.
We let Alice and Bob each choose with 50\thinspace \% probability one of
two settings, $a,b=0,1$, and obtain measurement results, $%
A,B=0,1$, respectively.
(Both sides in Ineq.~(\ref{CHSH}) are divided by 4 for later convenience.)

Assuming freedom of choice and realism, violations of the CHSH inequality imply that either OI or SI,
or both, needs to be given up. In the framework of `experimental metaphysics' \cite{Shim} it is violation
of the condition OI that is supposed to be responsible for the violation of the
CHSH inequality, and it is extensively argued by many philosophers of this school
that this is not an instance of `action at a distance' but only of some innocent `passion at a distance':
one passively comes to know the faraway outcome, but one cannot actively change it.
In contrast, violation of SI allows superluminal signaling at the hidden variable level,
as the distant observer can freely choose his measurement settings. Our analysis shows
that SI and OI do not provide us with the full picture of what (non-) local information is needed in violations of Bell inequalities.
Both SI and OI are in fact conditions on the conditional statistical independence of probabilities for the
\emph{local} outcome only. They do not exhaust all possibilities how information about \textit{distant} settings and outcomes can be inferred (locally or non-locally). Here we will provide such an analysis. We thus no longer focus on conditional statistical independence, but instead on the availability of (non-) local information.

We will show that within the framework of non-local realistic theories it is impossible
to model a violation of the CHSH inequality without having information in one laboratory
about \emph{both} the setting and the outcome at the distant one. Thus, the availability of
non-local information 
displayed by models that violate
the CHSH inequality is necessarily about both the non-local settings and the
outcomes, despite the fact that it is \emph{not necessary} that the models are
\emph{both} explicitly setting dependent ($\neg$ SI) \emph{and} outcome dependent ($\neg$ OI).

In order to prove our results we consider a local hidden-variable model augmented with
information available to Alice about Bob's laboratory. While it is not necessary, it is instructive to think about this information as
one-way classical communication from Bob to Alice. In every run of the
experiment, Alice and Bob first choose their settings ($a$ and $b$) and
receive hidden variables $\lambda $ which are independent of the choice of
the settings. Then, Bob (or some process in his lab) generates the outcome $%
B $ which in general depends on $\lambda $ and $b$. Next, Bob generates the
\textit{message} $\mathcal{X}$ which depends on $\lambda $, $b$ and $B$.
Both the generation of $B$ and of $\mathcal{X}$ are, in general,
probabilistic processes. It is assumed that the exact mechanism how $B$ and $%
\mathcal{X}$ are generated is known to Alice. Finally, $\mathcal{X}$ is
transmitted to Alice who uses her optimal strategy, based on the knowledge
of her setting $a$, the shared hidden variables $\lambda $, Bob's mechanisms and the message $%
\mathcal{X}$, to produce her outcome $A$ in order to maximally violate the
CHSH inequality.

From Alice's perspective, the CHSH inequality reads
\begin{align}
\frac{1}{2}P(A=B|a=0)+\frac{1}{2}P(A=B\oplus b|a=1)\leq \frac{3}{4},
\label{ALICE_CHSH}
\end{align}%
where, e.g.\ $P(A=B|a=k)$ is the probability that the outcome of Alice
equals that of Bob, given she has chosen the $k$th setting.
We shall show that the probabilities entering Ineq.~(\ref{ALICE_CHSH}) can be interpreted
as a measure of the information Alice has about Bob's measurement setting and
outcome. For this aim, we introduce the \textquotedblleft guessed
`information'\textquotedblright\ (GI)\footnote{This measure of information is related to conditional min-entropy. Min-entropy is defined by $H_{\min}(Y) = - \log (\max P(Y))$, and the conditional one reads $H_{\min}(Y|X) = \sum_x P(x) H_{\min}(Y|x) = - \sum_x P(x) \log (\max P(Y|x))$. This is just our definition \eqref{guess}, with the only difference being the $\log$ function. For this reason a case could be made to use the terminology \textquotedblleft guessed probability\textquotedblright\ rather than \textquotedblleft guessed
information\textquotedblright.}:
\begin{equation}\label{guess}
J(\mathcal{X}\rightarrow \mathcal{Y}):=\sum_{i}P(\mathcal{X}=i)\max_{j}[P(%
\mathcal{Y}=j|\mathcal{X}=i)],
\end{equation}%
where $\mathcal{X}$ takes values $i=1,...,X$ and $\mathcal{Y}$ values $%
j=1,...,Y$. The value of $J(\mathcal{X}\rightarrow \mathcal{Y})$ gives the
average probability to correctly guess $\mathcal{Y}$ knowing the value of $%
\mathcal{X}$. Its maximum is $1$ and then $\mathcal{Y}$ is fully specified by
$\mathcal{X}$. The minimum of $J(\mathcal{X}\rightarrow \mathcal{Y})$
equals $\frac{1}{Y}$ and then $\mathcal{X}$ reveals no information
about $\mathcal{Y}$. We note that GI reaches its minimum when the mutual
information is $I(\mathcal{X}:\mathcal{Y})=0$, and it is maximal when $I(%
\mathcal{X}:\mathcal{Y})=\log Y$. As an example, freedom of choice can be stated as  $J(\lambda \to a,b)=\frac{1}{4}$,
i.e., $\lambda$ cannot reveal any information about the settings $a$ and $b$.
This implies the weaker condition $J(\lambda \to b)=\frac{1}{2}$, which is sufficient for our analysis and to which we refer as freedom of choice throughout the rest of the paper.


Alice now uses an optimal maximization strategy so as to maximally violate the CHSH inequality.
Consider the case in which Alice chooses $a=0$. Her goal is to maximize the
probability $P(A=B | a=0)$ given the communicated value $\mathcal{X}$ and
the received hidden variables $\lambda$.
This maximized probability is just the average probability to correctly guess $B$ given $\mathcal{X}$ and $%
\lambda$: $J(\lambda, \mathcal{X} \to B)$. Similarly, if her setting is $a=1$%
, the maximal probability $P(A = B \oplus b | a = 1)$ equals $J(\lambda,%
\mathcal{X} \to B \oplus b)$. This allows to phrase the CHSH inequality in terms of the GI's
\begin{equation}  \label{iCHSH}
\frac{1}{2} J(\lambda,\mathcal{X} \to B) + \frac{1}{2} J(\lambda,\mathcal{X}
\to B \oplus b) \le \frac{3}{4}.
\end{equation}

In case Alice's scheme is not optimal $P(A=B|a=0)$ is upperbounded by
$J(\lambda,\mathcal{X}\rightarrow B)$, and likewise $P(A=B\oplus b|a=1)$
is upperbounded by $J(\lambda,\mathcal{X}\rightarrow B\oplus b)$.  Therefore,
even if Alice's strategy is not optimal violation of \eqref{iCHSH} is necessary for a
violation of the CHSH inequality.
This holds for any of the eight different CHSH inequalities.

We are now in the position to prove that a \textit{necessary} condition for the violation of Bell's
inequalities within non-local realism is that both information about the
setting and about the outcome produced at one lab must be available at the
distant lab. If there is no outcome information available,
i.e.,\ $J(\lambda , \mathcal{X}\rightarrow B)=\frac{1}{2}$,
the left-hand side of Ineq.~(\ref{iCHSH}) cannot exceed $\frac{3}{4}$.
To prove that setting information is
also necessary, note that if one knows both $B$ and $B\oplus b$, one also
knows $b$. Thus, the average probability of correctly guessing $b$ is
greater or equal to the product of the average probabilities for the correct
guess of $B$ and $B\oplus b$:
\begin{equation}
J(\lambda ,\mathcal{X}\rightarrow b)\geq J(\lambda ,\mathcal{X}\rightarrow
B)J(\lambda ,\mathcal{X}\rightarrow B\oplus b).  \label{jjj}
\end{equation}%
If $\mathcal{X}$ and $\lambda $ carry no information about the setting,
i.e., $J(\lambda ,\mathcal{X}\rightarrow b)=\frac{1}{2}$,
Ineq.~(\ref{jjj}) can be
rewritten as $J(\lambda ,\mathcal{X}\rightarrow B\oplus b)\leq \frac{1}{2}%
J^{-1}(\lambda ,\mathcal{X}\rightarrow B)$, which implies
\begin{eqnarray}
\frac{1}{2}J(\lambda ,\mathcal{X} &\rightarrow &B)+\frac{1}{2}J(\lambda ,%
\mathcal{X}\rightarrow B\oplus b)  \notag \\
&\leq &\frac{1}{2}J(\lambda ,\mathcal{X}\rightarrow B)+\frac{1}{4J(\lambda ,%
\mathcal{X}\rightarrow B)}.
\end{eqnarray}%
for the left-hand side of (\ref{iCHSH}). This value is less or equal $\frac{3%
}{4}$ for the whole range of $J(\lambda ,\mathcal{X}\rightarrow B)\in
\lbrack \frac{1}{2},1]$. Thus, if there is no setting information, the
violation of Ineq.~(\ref{iCHSH}), or Ineq.~(\ref{CHSH}), is
impossible.

Although both the information about the distant setting and about the
distant outcome must be available at the local laboratory to
have a violation, we show that, given freedom of choice, the information
about the distant setting has to be transmitted non-locally, whereas it is
possible that the information about the distant outcome can be obtained
without any transmission from the shared hidden variables.
This is shown by a further analysis of
what information has to be
transmitted via the message $\mathcal{X}$,
over and above the information in the hidden variable $\lambda$.
This also allows us to analyze the above mentioned asymmetry
between the outcome
and setting information in a more formal way.

To this end,
we introduce a measure of information,
that we call \textquotedblleft transmitted `information'\textquotedblright\
(TI), which is the difference of the averaged probability of correctly
guessing the value of the variable $\mathcal{Y}$ when knowing $\mathcal{X}$
and $\lambda $, and the one when knowing only $\lambda $:
\begin{equation}
\Delta _{\lambda }(\mathcal{X}\rightarrow \mathcal{Y}):=J(\lambda ,\mathcal{X}%
\rightarrow \mathcal{Y})-J(\lambda \rightarrow \mathcal{Y}).
\end{equation}%
$\Delta _{\lambda }(\mathcal{X}\rightarrow \mathcal{Y})$ takes values
between 0 and $1-\frac{1}{Y}$. Its lowest value means that transmission of $%
\mathcal{X}$ does not increase Alice's chances of guessing the correct value
of $\mathcal{Y}$;  $\mathcal{X}$ carries no \textit{new} information about $\mathcal{Y}$
that is not already available to Alice through $\lambda $.

We have already established that either  $J(\lambda ,\mathcal{X}%
\rightarrow B)=\frac{1}{2}$ or $J(\lambda ,\mathcal{X}\rightarrow b)=\frac{1%
}{2}$ implies no violation of the CHSH inequality. The asymmetry between the
outcome and setting information originates from the freedom of choice
assumption $J(\lambda \rightarrow b)=\frac{1}{2}$, which leads to
\begin{equation}
J(\lambda ,\mathcal{X}\rightarrow b)=\Delta _{\lambda }(\mathcal{X}%
\rightarrow b)+\frac{1}{2}.
\end{equation}%
We see that $\Delta _{\lambda }(\mathcal{X}\rightarrow b)=0$ leads to $%
J(\lambda ,\mathcal{X}\rightarrow b)=\frac{1}{2}$ which means no violation
of the CHSH inequality. On the other hand, there is no assumption
corresponding to freedom of choice regarding the outcomes, i.e.,\ there are no
physical grounds for assuming $J(\lambda \rightarrow B)=\frac{1}{2}$.
Instead, one has
\begin{equation}
J(\lambda ,\mathcal{X}\rightarrow B)=\Delta _{\lambda }(\mathcal{X}%
\rightarrow B)+J(\lambda \rightarrow B).
\end{equation}%
Thus, even if $\Delta _{\lambda }(\mathcal{X}\rightarrow B)=0$, it is
possible that $J(\lambda ,\mathcal{X}\rightarrow B)>\frac{1}{2}$, if $%
J(\lambda \rightarrow B)>\frac{1}{2}$. Also $J(\lambda \rightarrow B)=\frac{1%
}{2}$ does not mean $J(\lambda ,\mathcal{X}\rightarrow B)=\frac{1}{2}$ since
$\Delta _{\lambda }(\mathcal{X}\rightarrow B)$ can be greater than 0.
Summing up, neither $\Delta _{\lambda }(\mathcal{X}\rightarrow B)=0$ nor $%
J(\lambda \rightarrow B)=\frac{1}{2}$ individually implies no violation of
the CHSH inequality, although both of them together do. One can easily construct a toy model\footnote{A random binary hidden variable $\lambda=0,1$ is distributed to Alice and
Bob. Bob's result for setting $b$ is defined as $B = \lambda \oplus b$.
Next, he communicates his outcome, $\mathcal{X} = B$. The result of Alice is
given by $A = a(\mathcal{X} \oplus \lambda) \oplus \mathcal{X}$.
Thus $J(\lambda,\mathcal{X}\rightarrow B)=1$ and $J(\lambda\rightarrow B)=\frac{1}{2}$.
Clearly, $A \oplus B = ab$, and the CHSH inequality is maximally violated.
However, there is an \textit{intrinsic setting information} in this model as Alice can read the
setting of Bob from the data available to her, $b = \mathcal{X} \oplus
\lambda$; and thus $J(\lambda,\mathcal{X}\rightarrow b)=1$ as well.
This is what allows violation of the CHSH inequality.
 Note that this model is deterministic and thus obeys OI (and violates SI),
despite the fact that it is the outcome $B$ that is being communicated.}
where $J(\lambda \rightarrow B)=\frac{1}{2}$
and violation occurs because the TI is $\Delta _{\lambda }(\mathcal{X}%
\rightarrow B)=\frac{1}{2}$.
A toy model where $\Delta _{\lambda }(\mathcal{X%
}\rightarrow B)=0$ and violation occurs is presented later. All different
cases are presented in Table \ref{table}, where our technical results are also  contrasted to those in terms of the conditions OI and SI.
\begin{table}[tbp]
\begin{tabular}{|c|c|}
\hline
Condition & Violation of CHSH possible? \\ \hline\hline
$J(\lambda,\mathcal{X}\rightarrow b)=\frac{1}{2}$ & No \\[1pt] \hline
$J(\lambda,\mathcal{X} \rightarrow B)=\frac{1}{2}$& No  \\[1pt]\hline
\hline
$J(\lambda\rightarrow b)=\frac{1}{2}$ & ~~~~~~~~~~~~~~~Yes (`freedom') \\[1pt] \hline
$J(\lambda \rightarrow B)=\frac{1}{2}$ & ~Yes$^*$  \\[1pt] \hline
\hline
$\Delta_\lambda(\mathcal{X}\to b)=0$ & No  \\[1pt] \hline
$\Delta_\lambda(\mathcal{X}\to B)=0$ & ~Yes$^*$  \\[1pt] \hline
\hline
SI: $P(A|a,b,\lambda)=P(A|a,\lambda)$& ~~Yes$^{**}$ \\[1pt]
\hline
OI: $P(A|a,b,B,\lambda)=P(A|a,b,\lambda)$& ~~Yes$^{**}$ \\[1pt]\hline
\end{tabular}%
\caption{The possibility of violation of the CHSH inequality in a local
realistic model augmented with communication of $\mathcal{X}$ from Bob to
Alice.
$J(\lambda,\mathcal{X}\rightarrow b)$ and $J(\lambda,\mathcal{X} \rightarrow B)$
are the \textquotedblleft guessed informations\textquotedblright\ by Alice,
where $\protect\lambda $ denotes the hidden variables, and $b$ and $B$ are Bob's
setting and outcome, respectively. $\Delta _{\protect\lambda }(\mathcal{X}%
\rightarrow b)$ and $\Delta _{\protect\lambda }(\mathcal{X}\rightarrow B)$
denote the \textquotedblleft transmitted information\textquotedblright\
to Alice about Bob's setting and outcome, respectively, which is communicated via $%
\mathcal{X}$. (See main text for their definitions.) \textquotedblleft
No\textquotedblright\ in the right column means that the corresponding
condition has to be violated to allow violation of the CHSH inequality.
\textquotedblleft Yes\textquotedblright\ means that there are models which
satisfy the condition and violate the CHSH inequality.
The starred \textquotedblleft Yes$^*$\textquotedblright\
in rows 4 and 6 indicate that for a violation either one of these conditions can
hold, but not both. Similarly for the doubly-starred  \textquotedblleft Yes$^{**}$\textquotedblright\
in rows 7 and 8, where for completeness we have included the previously known
results in terms of SI and OI.}
\label{table}
\end{table}

To reinforce our conclusion that the freedom of choice assumption is responsible for
the asymmetry, consider the possibility of `superdeterminism', where everything is
determined by the hidden variables, even the setting choices. In that case both
$J(\lambda\rightarrow b)=1$ and $J(\lambda \rightarrow B)=1$, and
consequently we have both
$\Delta_\lambda(\mathcal{X}\to b)=0$  and $\Delta_\lambda(\mathcal{X}\to B)=0$:
In that case there simply is no new information to be transferred, i.e.
$\mathcal{X}$ is redundant as $\lambda$ determines all there is to know.
Settings and outcomes thus here appear on equal footing, and 
the conditions for violation
of the CHSH inequality become identical for both.  Indeed,
only by giving up superdeterminism and allowing for freedom of choice for the
settings we see the asymmetry between settings and outcomes arise.  The
assumption of freedom of choice of the settings enforces that, in order to get a
violation of the Bell inequality, the message $\mathcal{X}$ \emph{must} contain
information about the setting, either implicit or explicit (though note that SI can be satisfied).
It is however not needed that it carries information about the outcome.

Now we study explicit examples of non-local realistic models which violate
the CHSH inequality. In all of them the non-local information $\mathcal{X}$ is
information about the distant setting.

Consider the model of Toner and Bacon~\cite{TB}. One of the parties sends
the bit $\mathcal{X}=\pm 1$ which is given by
$\mathcal{X}=\mathrm{sgn}(\vec{b}\cdot \vec{\lambda}_{1})\,\mathrm{sgn}(\vec{b}\cdot \vec{\lambda}_{2})$,
where $\vec{b}$ is a unit Bloch vector corresponding to Bob's setting and $%
\vec{\lambda}_{1}$ and $\vec{\lambda}_{2}$ are also unit vectors which play
the role of hidden variables.
The communication in this model can be compressed
to $C\approx 0.85$ bits~\cite{TB}. Exactly the same value is
obtained for the mutual (Shannon) information between the bit sent and the setting, $I(%
\mathcal{X}:\vec{b})=C$.

The model of~\cite{TB} perfectly simulates all possible measurement results
obtained on the singlet state. If one aims at simulation of the maximal
violation of the CHSH inequality (with four fixed settings) allowed by
quantum mechanics, then less communication is needed as shown in the
following toy model with $\Delta _{\lambda }(\mathcal{X}\rightarrow B)=0$. A
binary random variable $\lambda =0,1$ is distributed from the source to
Alice and Bob. His outcome for any choice of the setting is defined as $%
B=\lambda $, implying $J(\lambda\rightarrow B)=1$ and thus
$J(\lambda, \mathcal{X}\rightarrow B)=1$ and  $\Delta _{\lambda }(\mathcal{X}\rightarrow B)=0$.
 If Bob's setting is $b=0$ he sends always $\mathcal{X}=0$, if
his setting is $b=1$ he sends $\mathcal{X}=1$ with probability $p=\sqrt{2}%
-1\approx 0.414$, and $\mathcal{X}=0$ otherwise. The outcome of Alice is
given by $A=a\mathcal{X}\oplus \lambda $.
In this model it is the information about the setting of Bob which is communicated:
the information content of $\mathcal{X}$ is $0.736$ bits and this is the mutual information $I(%
\mathcal{X}:\vec{b})$ between $\mathcal{X}$ and the setting of Bob.
Note that classical players
exchanging this amount of information achieve efficiency of quantum
solutions to communication complexity problems and games based on the CHSH
inequality~\cite{CCP}.

In the Leggett-type~\cite{LEGGETT} non-local model of Ref.~\cite{NL}, a real
unit vector, i.e.,\ an infinite number of bits, parameterizing the setting is
being sent from one party to another, thus  $\Delta _{\lambda }(\mathcal{X}\rightarrow b)>0$.
Note that this model violates SI, but obeys OI as it is deterministic \cite{Jarr}.

Our last example is Bohm's theory~\cite{BOHM}.  Although here there is no explicit
communication process, the specific dynamics of this theory allows for setting information to be non-locally available;
the information about the
setting of the apparatus in one lab enters the formula for the velocity of
the particle in the other lab. The analysis of the double Stern-Gerlach
experiment shows that the velocity of one of the particles is given by~\cite%
{ACP} $v_{1}=c_{1}\tanh (c_{2}\kappa )$, where $\kappa $ is a parameter that
describes the ratio between the magnetic field strengths at the two
distant laboratories. The constants $c_{1}$
and $c_{2}$ do not depend on $\kappa $. Therefore, 
the local measurement outcome and the knowledge about the velocity  \emph{would} allow to infer the distant setting.  
Since tanh is an injective function,
to determine $v_{1}$ all the bits defining $\kappa $ have to be known to the
mechanism that generates this velocity.

This last example shows that there need not be an actual communication process
and our results are valid  outside of the one-way communication paradigm.
Indeed, it is irrelevant for our results how Alice obtained the information $\mathcal{X}$;
one can
think of it as extra information that tells about Bob's situation, and which is
\emph{somehow} available to Alice.

Conclusions---
This work gives the general
conditions which every non-local hidden-variable theory has to satisfy in
order to allow for violation of the CHSH inequality. For there to be such a violation
it must be the case that information about \textit{both} the outcome and setting at one laboratory
is {\it available} at the distant one, despite the fact that there is \textit{no need} for both non-local
setting and outcome dependence in the conditional (hidden) probabilities. The role of the setting is shown to be
fundamentally different from that of the outcome and this asymmetry is
shown to be due to the assumption of the experimenter's free setting choice.
Because of this freedom the only way to learn a distant setting is to have non-local information transferral.
By contrast, it is possible that the distant outcome can also be learnt from the shared
hidden-variables, without any such non-local information transferral. The necessity that---within hidden variable models and freedom of choice---information about freely chosen distant settings has to be available in a space-like separated way, seriously questions the possibility of Lorentz-invariant completion of quantum mechanics.
This remark applies to both deterministic \cite{jones} as well as to stochastic models and clearly goes beyond what can be concluded on the basis of an analysis using only the conditions SI and OI.

We acknowledge support from the Austrian Science Foundation FWF within
Project No.\ P19570-N16, SFB-FoQuS and CoQuS No.\ W1210-N16, the European
Commission, Project QESSENCE, the National Research Foundation and
Ministry of Education in Singapore, and the Foundational Questions Institute
(FQXi). The collaboration is a part of an \"OAD/MNiSW program.

\end{document}